\documentclass[aps,prc,preprint,amsmath,showpacs,english]{revtex4}
\usepackage[T1]{fontenc}
\usepackage[latin1]{inputenc}
\usepackage{float}
\usepackage[dvips]{graphicx}
\usepackage{babel}
\makeatother

\begin{document}

\title{Truncated Moments of Structure Functions}
%\footnote{Talk given at GRC Photonuclear Reactions, Tilton School, August 10 - 15,   
%                 2008.}}

\author{A. Psaker}
\affiliation{American University of Nigeria, Yola, Nigeria\\}

\begin{abstract}
We present a novel new approach to study quark-hadron duality using truncated moments of 
structure functions, and determine the degree to which individual resonance regions are 
dominated by leading twist.
\end{abstract}


\maketitle

%%%%%%%%%%%%%%%%%%%%%%%%%%%%%%%%%%%%%%%%%%%%%%%%%%
\section{Introduction}

In physics of strong nuclear interactions, the observed hadrons at low energies 
can be described in terms of meson and baryon degrees of freedom, while at high 
energies the description is formulated in terms of quarks and gluons. Quark-hadron 
duality is an intriguing phenomenon which, by connecting the two energy regimes, 
provides a dual description of hadronic variables. Its understanding, however, still 
represents a fundamental challenge. 

In Ref.~\cite{Psaker} we present a novel new 
approach to study quark-hadron duality using truncated moments of structure functions, 
or the integrals of structure functions over restricted regions of Bjorken variable $x$. 
In particular, by studying the $Q^2$ evolution of structure functions integrated over 
specific nucleon resonance regions, where $Q^2$ is the virtuality of the exchanged 
photon, we determine the degree to which individual resonance regions are dominated 
by leading twist, and hence quantify their higher twist content.

The phenomenon of quark-hadron duality was first observed in inclusive 
electron-nucleon scattering at SLAC by Bloom and Gilman in 1970  \cite{BG}. 
It reflects the similarity between structure functions averaged over the resonance region 
and the scaling, or leading twist function. The former is characterized by hadronic 
bound states, whereas the latter describes the high energy deep inelastic continuum, 
in other words, scattering from free quarks \cite{MEK}. The oscillation of nucleon resonances 
around the scaling curve shows that hadrons follow the QCD scaling behavior. The only 
rigorous interpretation of the so-called Bloom-Gilman duality within the theoretical framework 
has been done in terms of the moments of structure functions. They were analyzed within the 
QCD operator product (or twist) expansion. Here the leading term in $1/Q^2$ expansion is 
given by matrix elements of leading twist quark-gluon bilocal operators, and is associated 
with free quark scattering. Furthermore, higher terms in the expansion correspond to 
nonperturbative (higher twist) quark-gluon interactions. In this language duality is explained 
as the suppression of higher twist contributions to the moments \cite{DGP}. 

Experimental data from recent years \cite{Niculescu} suggest that quark-hadron duality not only 
exists over the whole resonance region but also locally. It holds over restricted regions of hadronic 
final state mass $W$, or even for individual resonances. Nevertheless, any insight about workings 
of local duality has been confined to QCD-inspired models of the nucleon. As such our 
understanding of quark-hadron duality in nucleon structure functions within QCD is incomplete.

Many earlier analyses of duality within a QCD context have used full moments of structure 
functions and hence quantified the higher twist content over all region of $x$. Our approach, 
on the other hand, allows one to study for the first time the distribution of higher twist corrections 
over various regions in $x$ (or $W$) in a well-defined systematic way.

%%%%%%%%%%%%%%%%%%%%%%%%%%%%%%%%%%%%%%%%%%%%%%%%%%
\section{Truncated Moments and Evolution}

In analogy with full moments, e.g. of a parton distribution function (PDF) 
$q(x,Q^2)$,
\begin{eqnarray}
{\cal M}_n(Q^2) & = & 
\int_0^1 dx\; x^{n-1}\ q(x,Q^2)\ ,
\label{eq:fullmoment}
\end{eqnarray}
truncated moments are defined as:
\begin{eqnarray}
{\cal M}_n(x_0,1,Q^2) & = & 
\int_{x_0}^1 dx\; x^{n-1}\ q(x,Q^2)\ ,
\label{eq:truncatedmoment}
\end{eqnarray}
where the integration over the Bjorken $x$ variable is restricted to $x_0 \leq x \leq 1$.  By truncating the integration region to some minimum value $x_0$, one avoids the problem of extrapolating parton distributions into unmeasured regions at small $x$. Several years ago this original idea was introduced to study structure function moments for which small-$x$ data were not available \cite{Forte}. Note that 
$x$ is related to the invariant mass squared $W^{2}$ of the virtual photon-hadron system through $W^2=M^2+Q^2\left(1-x\right)/x$, where $M$ is the nucleon mass. Thus an infinite energy is required 
to reach the $x \to 0$ limit, and accordingly some extrapolation to $x=0$ is always needed in practice to evaluate the moment. Unlike full moments, truncated moments satisfy the evolution equations, which mix the lower moments with the higher ones. 

An alternative approach, the so-called diagonal formulation of truncated moments, was later developed to avoid the complication of mixing \cite{Kotlorz}. It was shown that truncated moments obey the DGLAP evolution with a modified splitting function in the Mellin convolution. Namely,
\begin{eqnarray}
\frac{d{\cal M}_n(x_0,1,Q^2)}{dt}
& = & \frac{\alpha_S(Q^2)}{2\pi}
\left( P'_n \otimes {\cal M}_n \right)(x_0,Q^2)\ ,
\label{eq:evolutionequationtruncatedmoments3}
\end{eqnarray}
where
\begin{eqnarray}
P'_n(z,\alpha_S(Q^2))
& = & z^n P(z,\alpha_S(Q^2))\ 
\label{eq:pprime1}
\end{eqnarray}
plays the role of the splitting function for the truncated moments. Moreover, by 
introducing the doubly-truncated moment of $q(x,Q^2)$, 
\begin{eqnarray}
{\cal M}_n(x_{\rm min},x_{\rm max},Q^2) & = & 
\int_{x_{\rm min}}^{x_{\rm max}} dx\; 
x^{n-1}\ q(x,Q^2)\ ,
\label{eq:doublytruncatedmoment}
\end{eqnarray}
the evolution equations for truncated moments can be generalized to any subset in the 
$x$-region, $x_{\rm min} \leq x \leq x_{\rm max}$.

In our work we partially follow the latter approach and apply it to the study of the proton 
structure function $F_2^p$ in the large-$x$ region, populated by nucleon resonances. 
We use recent high-precision data from Jefferson Lab experiment E91-110 \cite{F2data}, 
and quantify the size of the higher twists for the lowest three moments in various regions of 
$W < 2$~GeV. By dividing the data into the three traditional resonance regions we extract the 
leading and higher twist content of each region.

%%%%%%%%%%%%%%%%%%%%%%%%%%%%%%%%%%%%%%%%%%%%%%%%%%
\section{Data Analysis and Extraction of Higher Twists}

In order to determine the degree to which nucleon structure function data in specific 
regions in $x$ (or $W$) are dominated by leading twist we first evolve the structure 
functions by brute-force using a suitable numerical integration routine, apply the 
target mass corrections (TMCs), and finally calculate their corresponding truncated 
moments over the integration range $W_{\rm th} \leq W \leq W_{\rm max}$, where 
$W_{\rm th} = M + m_\pi$ is the inelastic threshold. Deviations of the evolved moments, 
computed to next-to-leading order (NLO) accuracy, from the experimental data at the new $Q^2$ 
then reveal any higher twist contributions in the original data. In particular, we analyze recent 
data on $F_2^p$  from Jefferson Lab covering a range in $Q^2$ from $\lesssim 1$~GeV$^2$ to 
$\approx$~6~GeV$^2$.

\begin{figure}[t]
\begin{center}
\includegraphics[scale=0.44]{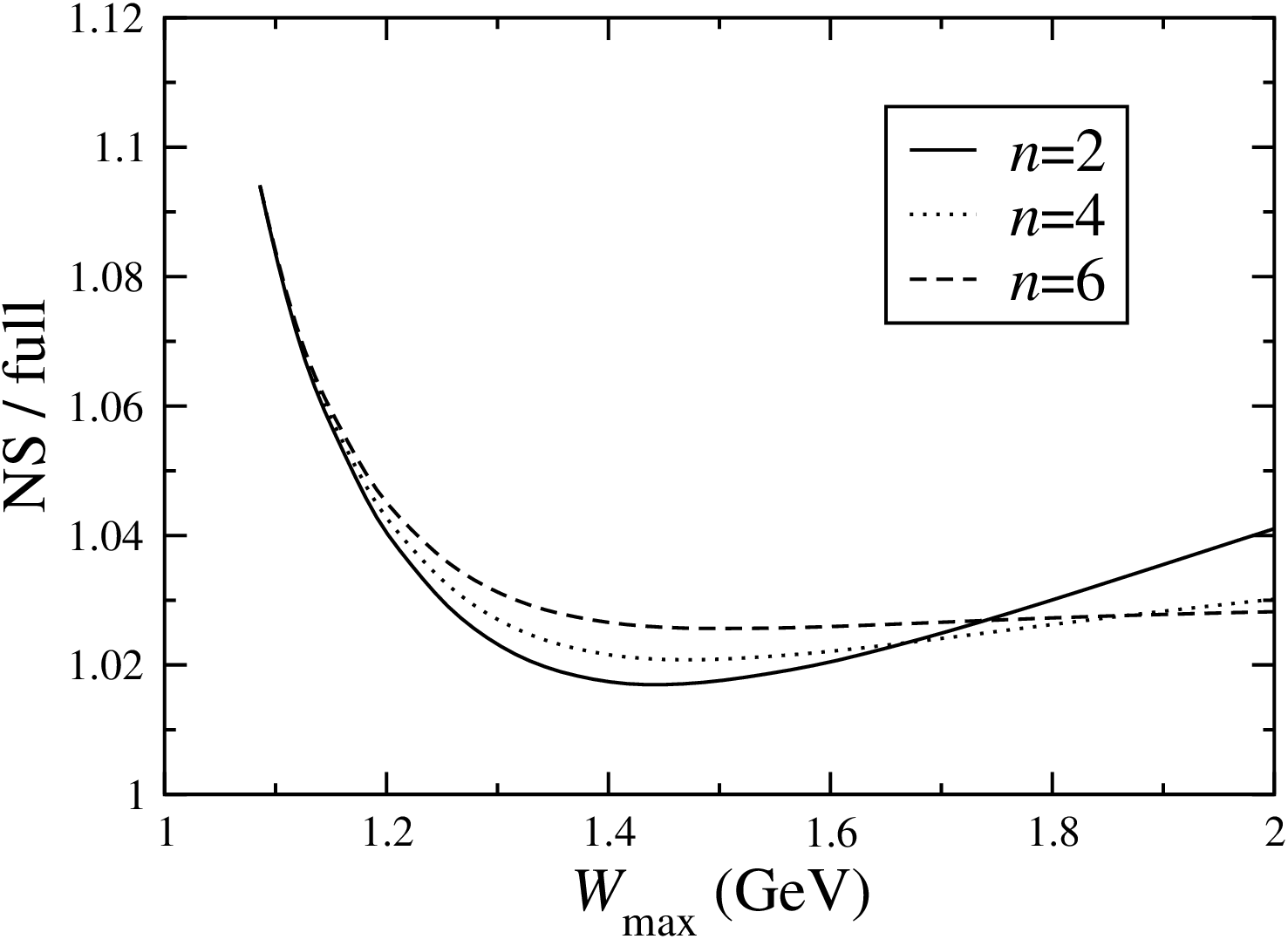}
\end{center}
\caption{Ratio of the truncated moments of $F_2^p$ evolved from
	$Q^2 = 25$ to 1~GeV$^2$, using NS and full evolution,
	versus the truncation point $W_{\rm max}$ for the $n=2$ (solid), 
	4 (dotted) and 6 (dashed) moments.}
\label{fig:1}
\end{figure}

The evolution of the measured truncated moments requires the structure function to 
be decomposed into its nonsinglet and singlet components, which are a priori unknown. 
We shall assume that in our region of interest (i.e. at moderate to large $x$) 
the structure function $F_2^p$ is well approximated by its nonsinglet component, and 
hence will evolve the truncated moments as nonsinglets. To test the accuracy of the 
nonsinglet (NS) evolution versus the full or exact evolution we use the trial 
function, whose decomposition into its nonsinglet and singlet components is known. 
In our case this is the leading twist proton $F_2$ structure function computed from 
the MRST2004 PDF fit \cite{MRST2004}. First we evolve its nonsinglet and singlet
components separately and then construct the full function, and secondly we evolve 
the trial function under the assumption that the total function can be treated as a 
nonsinglet. The comparison between the truncated moments calculated from the two 
evolutions gives the accuracy of the NS evolution of the various moments 
with respect to the exact results using both singlet and nonsinglet evolution. The 
ratios of these are plotted in Fig.~\ref{fig:1} as a function of the truncation 
point $W_{\rm max}^2 = M^2 + Q^2 (1/x_{\rm min}-1)$. In the traditional nucleon 
resonance region, $1.2 \lesssim W_{\rm max} \lesssim 2$~GeV, the differences 
between the full and NS evolution are of the order 2--4\% and increase with increasing 
$W_{\rm max}$. The $n=2$ moment is most sensitive to singlet evolution. As expected, 
the differences are smaller for the higher moments. Thus in the region 
relevant for our study one can therefore conclude that the evolution error, or the 
uncertainty introduced by evolving the $F_2^p$ truncated moments as nonsinglets, is less than 
4\%, which will be included in the errors in our final results.

\begin{figure}[t]
\begin{center}
\includegraphics[scale=0.45]{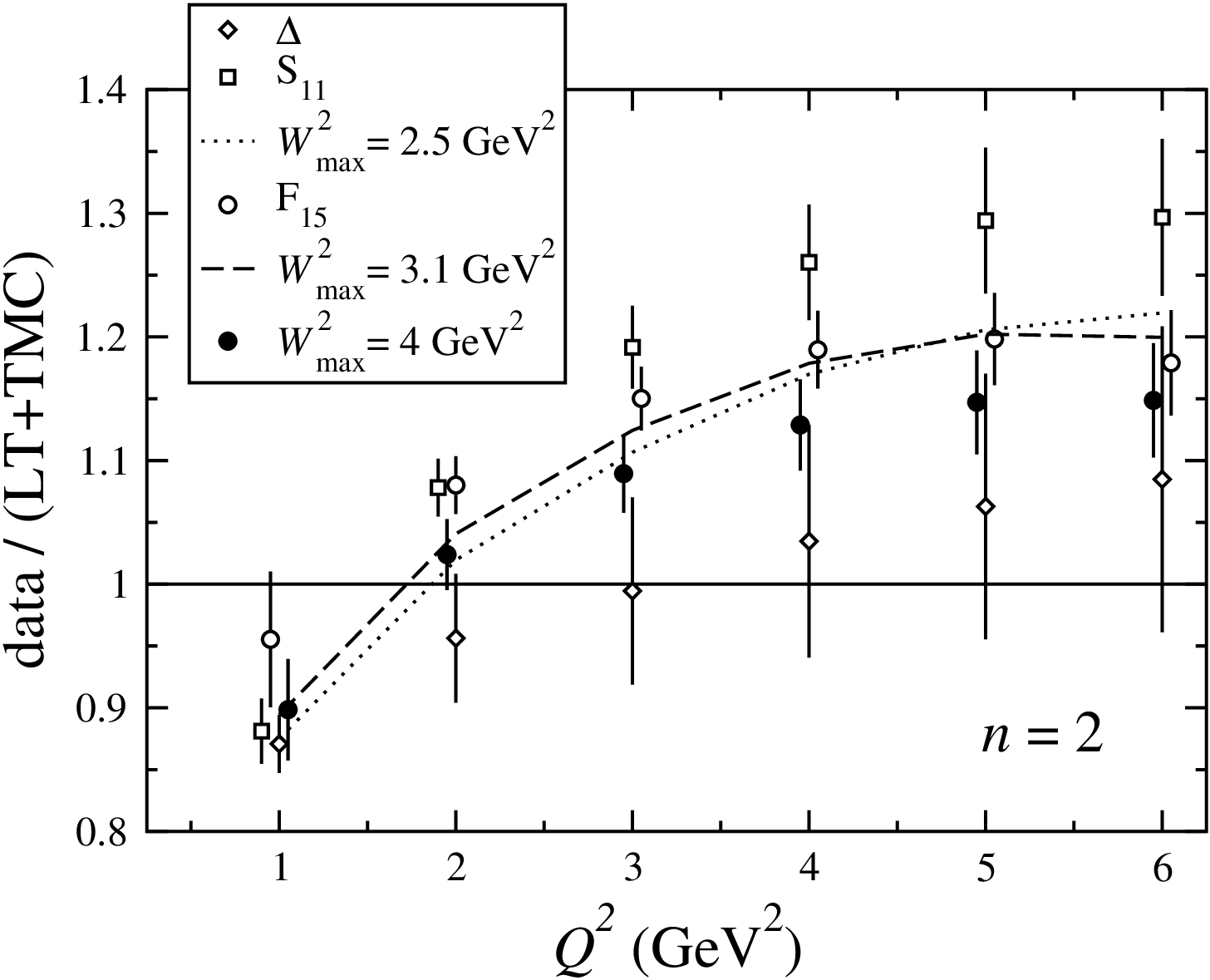}
\end{center}
\caption{$Q^2$ dependence of the ratio of $n=2$ truncated moments
	${\cal M}_2$ calculated from the data and from leading twist
	evolution from $Q_0^2 = 25$~GeV$^2$ (including TMCs), for
	various intervals in $W$: the first ($\Delta$) resonance region
	(diamonds), second ($S_{11}$) resonance region (squares),
	the first and second combined, corresponding to
	$W_{\rm max}^2 = 2.5$~GeV$^2$ (dotted curve), third ($F_{15}$)
	resonance region (open circles), first three regions combined,
	$W_{\rm max}^2 = 3.1$~GeV$^2$ (dashed curve), and the entire
	resonance region $W_{\rm max}^2 = 4$~GeV$^2$ (filled circles).
	Note that some of the points are offset slightly for clarity.}
\label{fig:4}
\end{figure}

In the analysis we assume that the $F_2^p$ data beyond some large value of $Q^2$ (in our case $Q^2=25$~GeV$^2$) are dominated by twist-2 contributions, which is consistent with most global analyses of PDFs. For example, a comparison of the DIS data fit shows the excellent agreement between the leading twist structure function, after correcting for TMCs, and the data at this scale. Moreover, we observe that the inclusion of TMCs has to be properly accounted for before drawing any conclusions about higher twists from data. To quantify the higher twist content of the specific resonance regions, and at different values of $Q^2$, we consider several intervals in $W$: $W_{\rm th}^2 \leq W^2 \leq$~1.9~GeV$^2$, corresponding to the traditional $\Delta(1232)$ (or first) resonance region; 
$1.9 \leq W^2 \leq 2.5$~GeV$^2$ for the $S_{11}(1535)$ (or second) resonance region; 
and $2.5 \leq W^2 \leq 3.1$~GeV$^2$ for the $F_{15}(1680)$ (or third) resonance region. 
The $n=2$ truncated moments corresponding to these regions are plotted 
in Fig.~\ref{fig:4} for various $Q^2$ values. It's worth noting that below $Q^2 = 1$~GeV$^2$ 
the applicability of a pQCD analysis becomes doubtful and the decomposition into leading 
and higher twists is no longer reliable.

\begin{figure}[t]
\begin{center}
\includegraphics[scale=0.45]{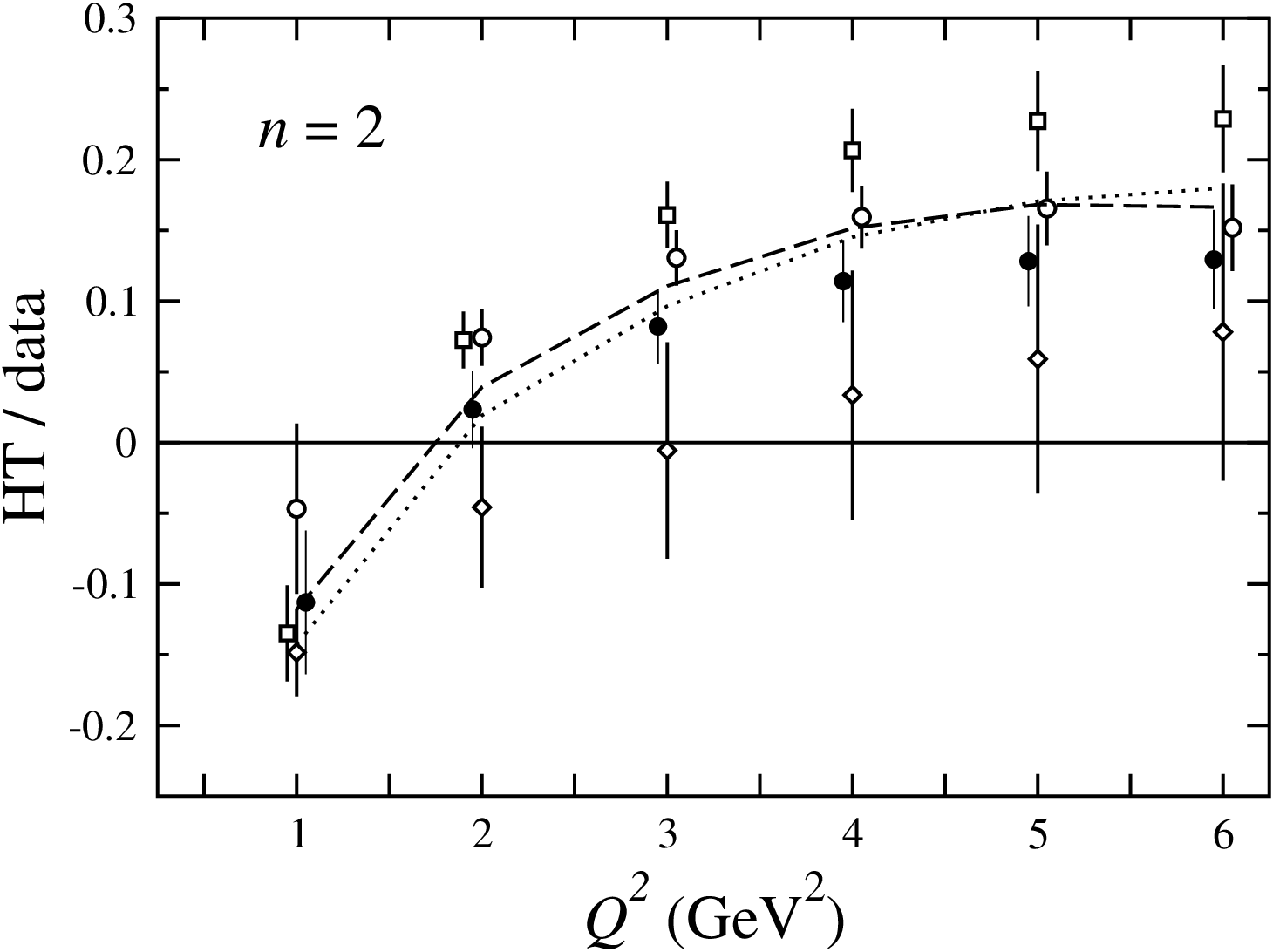}
\end{center}
\caption{$Q^2$ dependence of the fractional higher twist (HT) 
	contribution to the $n=2$ truncated moment data, for various 
	intervals in $W$ (as in Fig.~\ref{fig:4}).}
\label{fig:5}
\end{figure}

The entire resonance region data (filled circles in Fig.~\ref{fig:4}) deviate from 
leading twist behavior at the level of $\lesssim 15\%$ for all values of $Q^2$ with a 
significant $Q^2$ dependence for $Q^2 \lesssim 3$~GeV$^2$, which is made more explicit 
in Fig.~\ref{fig:5}. Here the higher twist contributions to ${\cal M}_2$, defined as 
the difference between the total and leading twist moments, are shown as ratios of the 
moments evaluated from the data. The strong $Q^2$ dependence of the higher twists is 
evident in the change of sign around $Q^2 = 2$~GeV$^2$, with the higher twists
going from $\approx -10\%$ at $Q^2 = 1$~GeV$^2$ to $\approx 10$--15\%
for $Q^2 \approx 5$~GeV$^2$. The higher twists, as expected, decrease at larger $Q^2$ 
once the leading twist component of the moments begins to dominate. 

Looking at the individual resonance regions we see that in the $\Delta$ region (diamonds) the higher 
twist contributions are smallest in magnitude at large $Q^2$, while they are largest for 
the $S_{11}$ region (squares). Combined, the higher twist contribution from the first 
two resonance regions (dotted curve) is $\lesssim 15\%$ in magnitude for all $Q^2$. 
Furthermore, the higher twist content of the $F_{15}$ region (open circles) is similar 
to the $S_{11}$ at low $Q^2$, however, it decreases more rapidly for $Q^2 > 3$~GeV$^2$. 
The higher twist content of the first three resonance regions combined (dashed curve) 
is $\lesssim 15$--20\% in magnitude for $Q^2 \leq 6$~GeV$^2$. Finally, integrating up 
to $W_{\rm max}^2 = 4$~GeV$^2$ (filled circles), the data on the $n=2$ truncated moment 
are found to be leading twist dominated at the level of 85--90\% over the entire 
$Q^2$ range.

The results in Figs.~\ref{fig:4} and \ref{fig:5} contain the experimental uncertainty on the $F_2$ 
data (statistical and systematic), and the uncertainty from the NS evolution of the data. For the 
experimental error we take an overall uncertainty of 2\% for all truncated moment 
data, except for the $n=4$ and $n=6$ moments at $W_{\rm max}^2 = 1.9$ and 4~GeV$^2$. 
Here the experimental uncertainties are 4\% and 3\% for ${\cal M}_4$, and 5\% and 
4.5\% for ${\cal M}_6$. The evolution error, on the other hand, is estimated by 
comparing the NS evolution with the full evolution using the MRST2004 fit, as in 
Fig.~\ref{fig:1}, with the appropriate correction factor applied at each $Q^2$ and 
$W$ interval.

\begin{figure}[t]
\begin{center}
\includegraphics[scale=0.32]{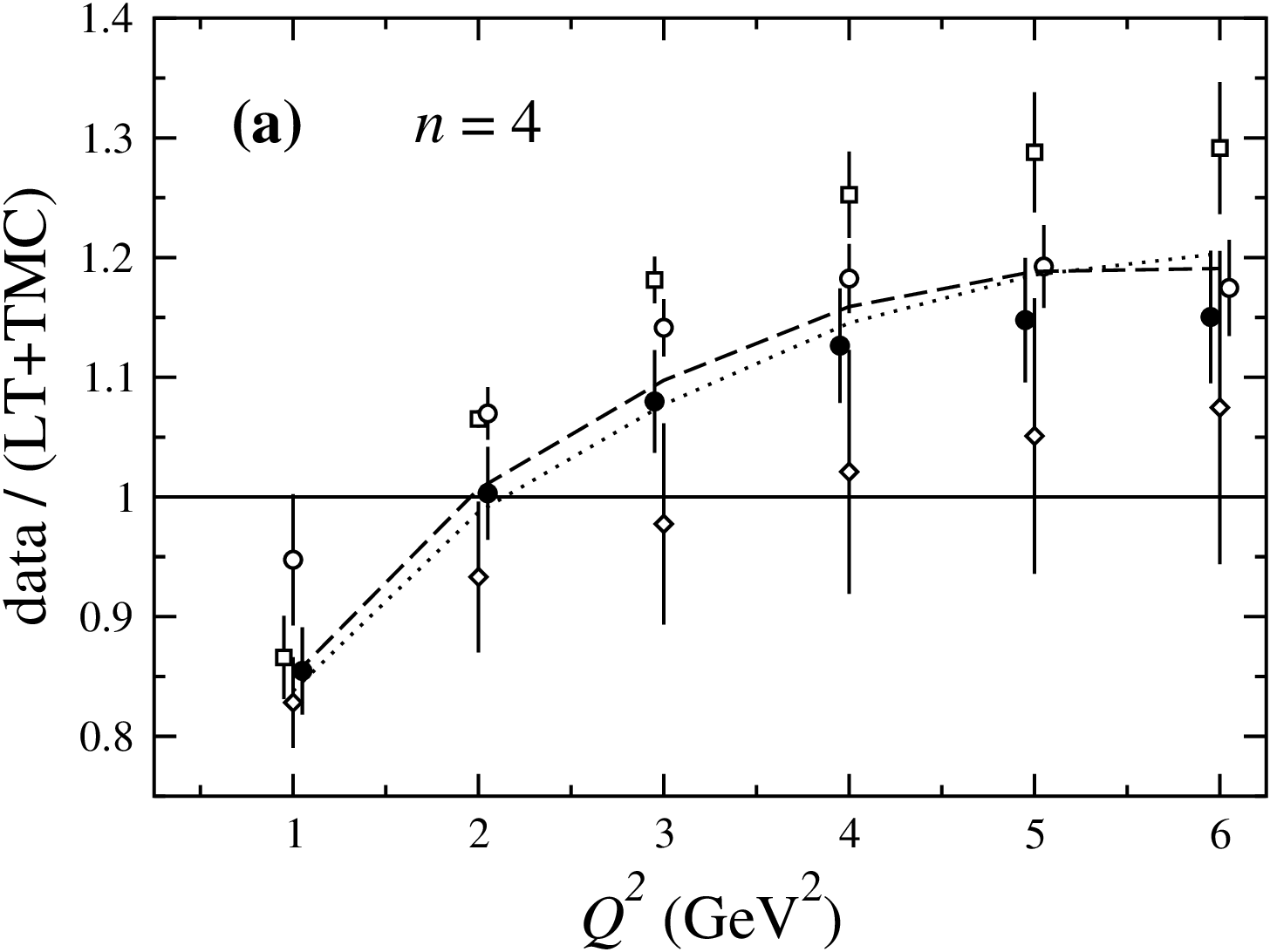}\ \
\includegraphics[scale=0.32]{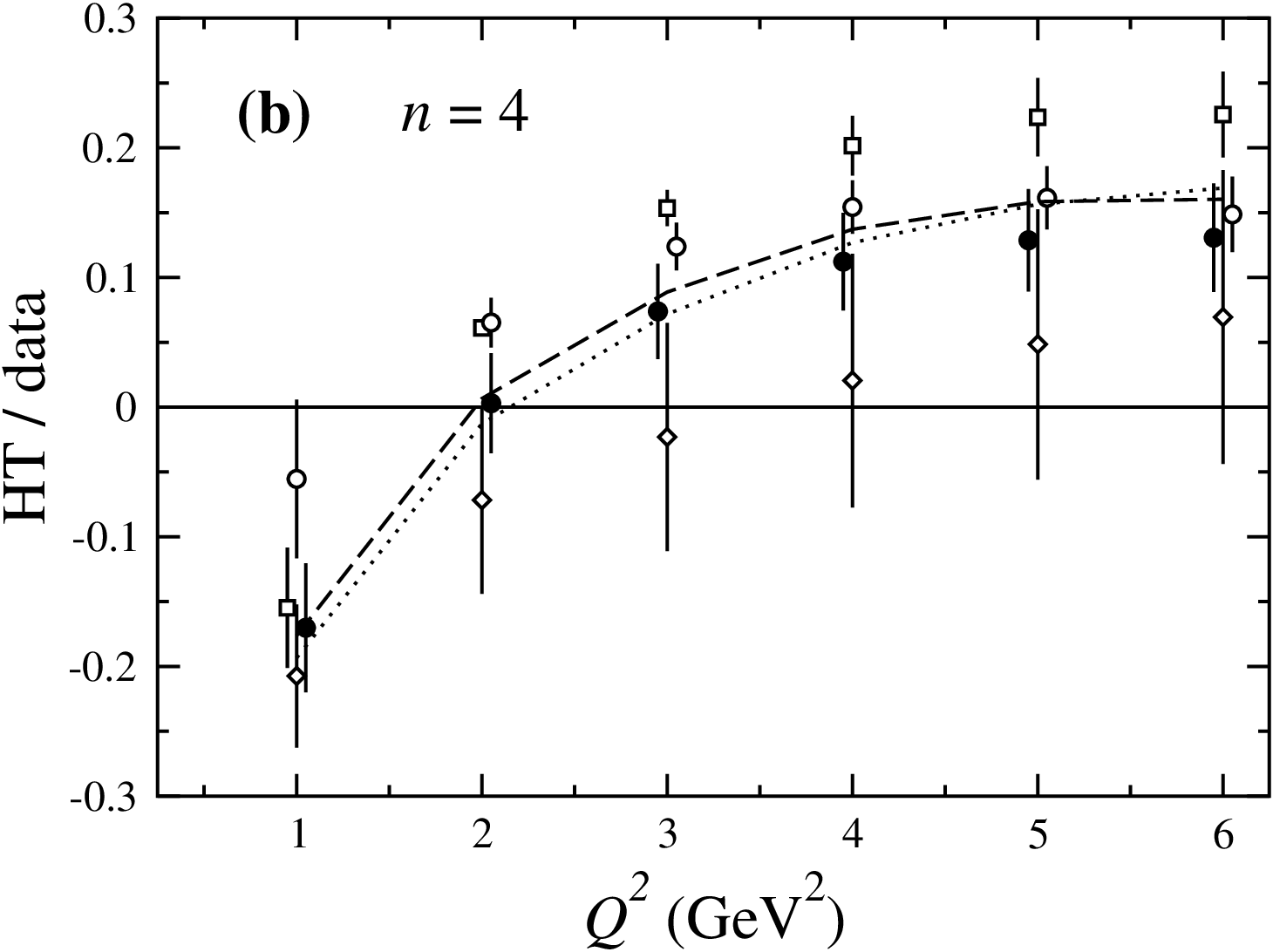}
\end{center}
\caption{(a) $Q^2$ dependence of the ratio of truncated moments
        ${\cal M}_4$ calculated from the data and from leading twist
        evolution from $Q_0^2 = 25$~GeV$^2$ (including TMCs), for
	various intervals in $W$ (labels as in Fig.~\ref{fig:4}).
	(b) Fractional higher twist contribution to the $n=4$
	truncated moment data, for various intervals in $W$
        (as in Fig.~\ref{fig:5}). \\ }
\label{fig:6}
\end{figure}

\begin{figure}[t]
\begin{center}
\includegraphics[scale=0.32]{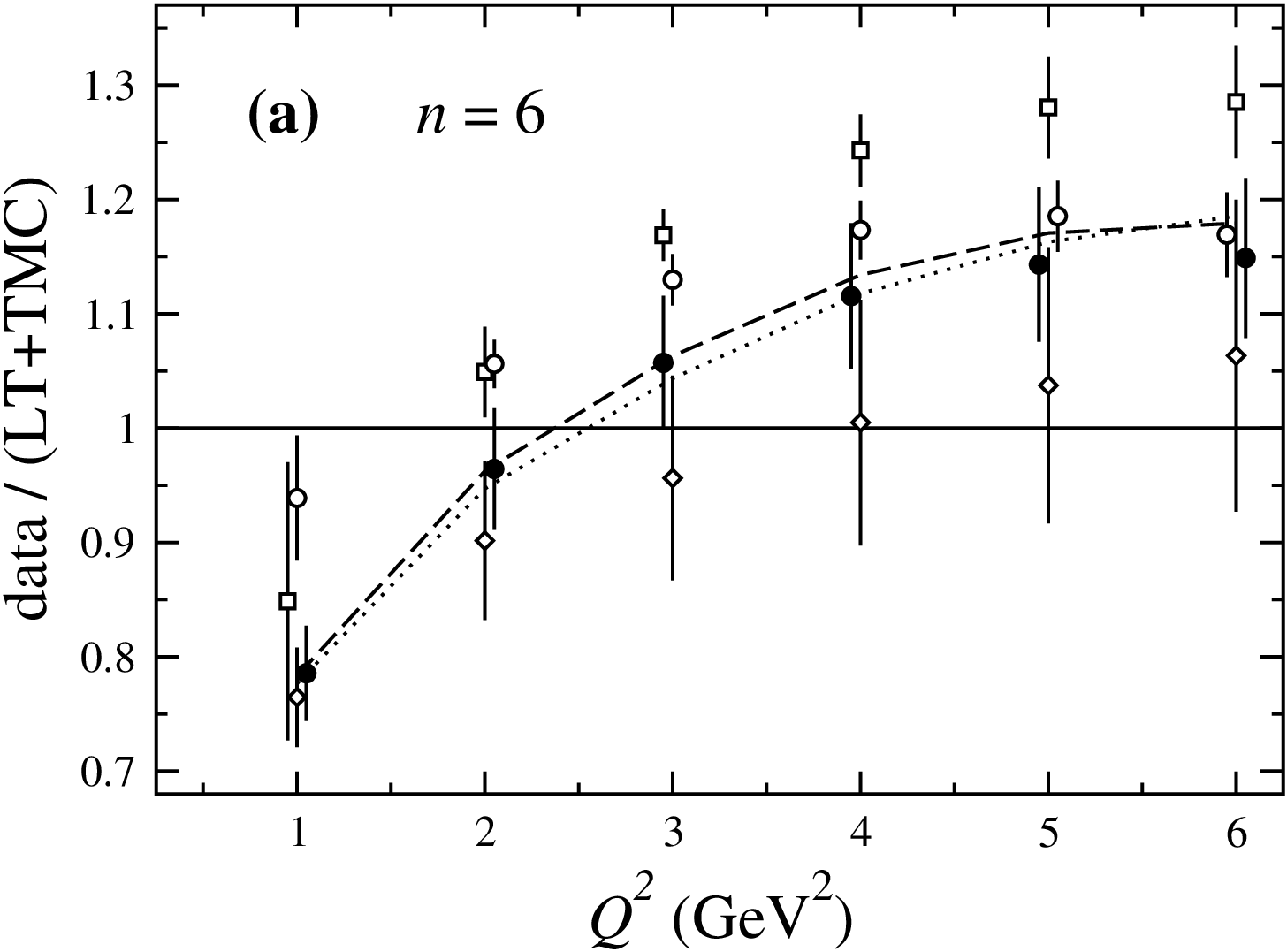}\ \
\includegraphics[scale=0.32]{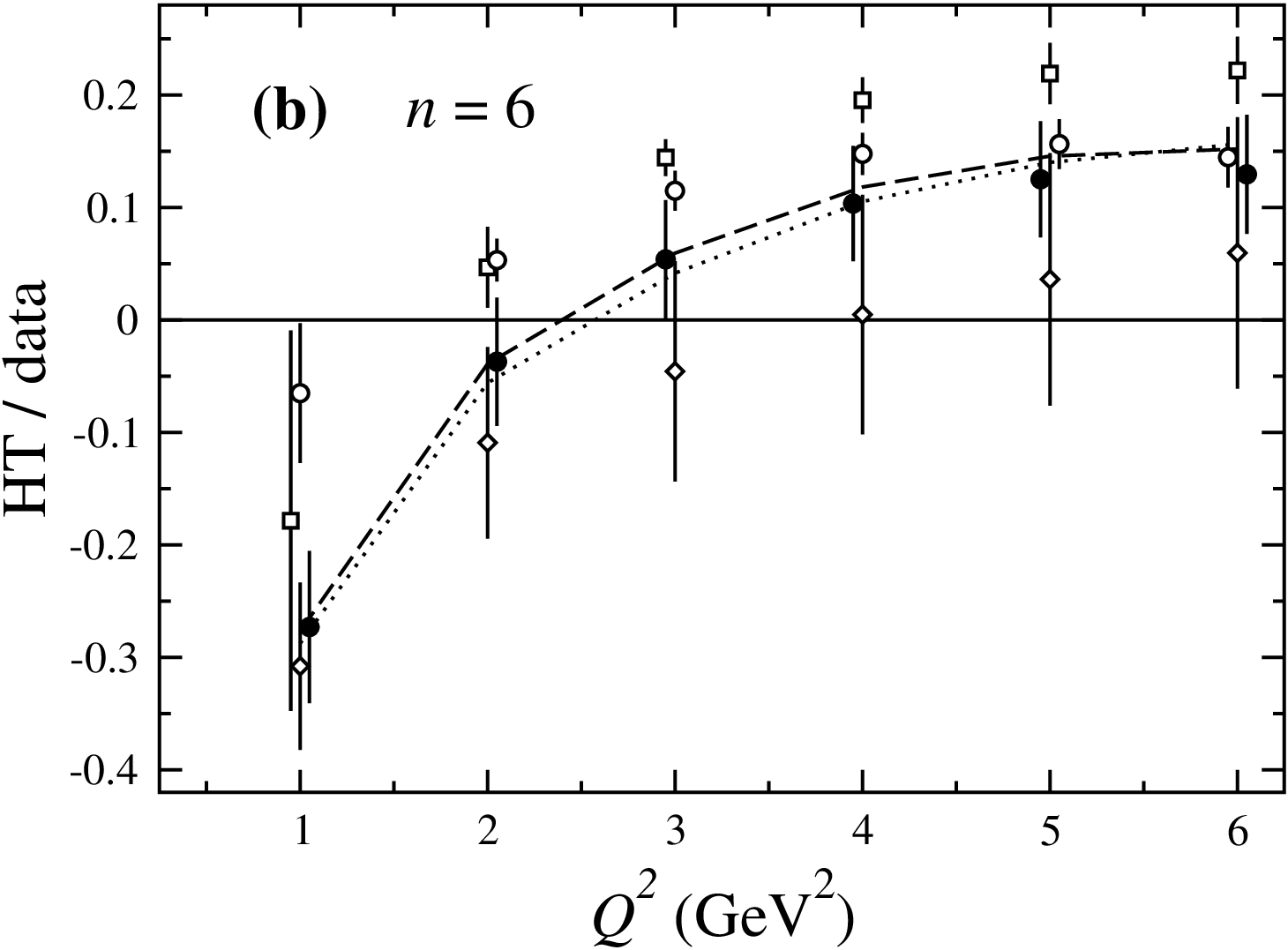}
\end{center}
\caption{(a) Ratio of truncated moments ${\cal M}_6$ calculated from the
	data and from leading twist evolution from $Q_0^2 = 25$~GeV$^2$ 
	(including TMCs), for various intervals in $W$ (labels as in 
	Fig.~\ref{fig:4}).
	(b) Fractional higher twist contribution to the $n=6$
	truncated moment data, for various intervals in $W$
	(as in Fig.~\ref{fig:5}).}
\label{fig:7}
\end{figure}

For the higher moments, illustrated in Figs.~\ref{fig:6} and \ref{fig:7}, the overall 
magnitude of the higher twists is qualitatively similar to the $n=2$ moments, although 
the $Q^2$ values at which they start decreasing in importance are larger. In addition, 
the higher twist contributions at low $Q^2$ are relatively larger for higher moments. 
For example, at $Q^2 = 1$~GeV$^2$ the magnitude of the higher twist component of the 
$W^2 < 4$~GeV$^2$ region increases from $\sim 10\%$ for the $n=2$ moment, 
to $\sim 15$--20\% for $n=4$, and $\sim 20$--30\% for $n=6$. This behavior can be 
understood from the relatively greater role played by the nucleon resonances in the 
large-$x$ region, which is emphasized more by the ($x$-weighted) higher moments.

%%%%%%%%%%%%%%%%%%%%%%%%%%%%%%%%%%%%%%%%%%%%%%%%%%%%%%
\section{Summary and Outlook}

Truncated moments of structure functions provide a firm foundation for the 
quantitative study of local quark-hadron duality within a perturbative QCD context. 
In this analysis we have precisely determined the degree to which duality holds as a function of $Q^2$ in various resonance regions (i.e. the degree to which individual resonance regions are dominated by leading twist), and have calculated for the first time their higher twist content. In case of the $n=2$ truncated moment of the proton structure function $F_2$ we find that deviations from leading twist behavior in the resonance region ($W \leq 2$~GeV) are at the level of $\lesssim 15\%$ for $Q^2 > 1$~GeV$^2$. Morevover, there is a significant $Q^2$ dependence in the ratio of moments of data to leading twist for $Q^2 \lesssim 3$~GeV$^2$, with the higher twists changing sign around $Q^2 = 2$~GeV$^2$. Our results indicate that in the second $(S_{11})$ and third $(F_{15})$ resonance regions the higher twist component is larger in magnitude than in the $\Delta$ resonance region. Similar behavior to the $n=2$ truncated moments is found for the $n=4$ and $n=6$ truncated moments, however, here due to the large-$x$ enhancement the resonances play relatively greater role, which then leads to larger higher twists at the same $Q^2$. In addition, we have also quantified the uncertainty associated with evolving the structure function data as a nonsinglet, which was found to be $\lesssim 4\%$.  

The illustrated analysis is an encouraging new approach to understanding local 
Bloom-Gilman duality within a well-defined theoretical framework. It opens the way 
to further study of local duality in other structure functions, such as the 
longitudinal structure function $F_L$ or spin-dependent structure functions.

%%%%%%%%%%%%%%%%%%%%%%%%%%%%%%%%%%%%%%%%%%%%%%%%%%%
\begin{acknowledgments}
I would like to thank Wally Melnitchouk, Cynthia Keppel and Eric Christy for their collaboration on the truncated moment analysis presented here. 
\end{acknowledgments}

%%%%%%%%%%%%%%%%%%%%%%%%%%%%%%%%%%%%%%%%%%%%%%%%%%%%

%%%%%%%%%%%%%%%%%%%%%%%%%%%%%%%%%%%%%%%%%%%
\end{document}